\documentclass[usenatbib]{mn2e}
\usepackage{graphicx}
\usepackage{ifthen}

\def\ltsima{$\; \buildrel < \over \sim \;$}
\def\simlt{\lower.5ex\hbox{\ltsima}}
\def\gtsima{$\; \buildrel > \over \sim \;$}
\def\simgt{\lower.5ex\hbox{\gtsima}}
\def\kpc{{\rm\,kpc}}

\def\msun{{\rm\,M_\odot}}

\def\pc{{\rm\,pc}}

\def\UseFigs{1}

\def\s{\ifmmode \widetilde \else \~\fi}
\def\={\overline}

\def\spose#1{\hbox to 0pt{#1\hss}}

\def\lta{\mathrel{\spose{\lower 3pt\hbox{$\mathchar"218$}}
     \raise 2.0pt\hbox{$\mathchar"13C$}}}
\def\gta{\mathrel{\spose{\lower 3pt\hbox{$\mathchar"218$}}
     \raise 2.0pt\hbox{$\mathchar"13E$}}}
\def\Dt{\spose{\raise 1.5ex\hbox{\hskip3pt$\mathchar"201$}}}    
\def\dt{\spose{\raise 1.0ex\hbox{\hskip2pt$\mathchar"201$}}}    

\def\dotsfill{\leaders\hbox to 1em{\hss.\hss}\hfill}

\def\Gyr{{\rm\,Gyr}}

\loadboldmathitalic 
\title[Sub-structure of the outer Galactic Halo from the 2~Micron All Sky Survey]
{Sub-structure of the outer Galactic Halo from the 2~Micron All Sky 
Survey\thanks{This publication  makes use of  data products from the  Two Micron
All Sky Survey, which is a  joint project of the University of Massachusetts
and  the Infrared  Processing  and Analysis  Center/California Institute  of
Technology, funded by the  National Aeronautics and Space Administration and
the National Science Foundation.}}

\author[R. A. Ibata, G. F. Lewis, M. J. Irwin \& L. Cambr\'esy]
{R. A. Ibata$^{1}$, G. F. Lewis$^{2}$, M. J. Irwin$^{3}$ \&
L. Cambr\'esy$^{4}$ \\
$^{1}$
Observatoire de Strasbourg, 11, rue de l'Universit\'e, F-67000, Strasbourg, 
France\\
$^{2}$
Anglo-Australian Observatory, P.O. Box 296, Epping, NSW 1710, Australia\\
$^{3}$
Institute of Astronomy, Madingley Road, Cambridge, CB3 0HA, U.K.\\
$^{4}$
IPAC, California Institute of Technology, Mail Code 100-22, Pasadena, CA
91125, USA}

\date{\today}

\begin{document} 

\maketitle 

\begin{abstract}
A pole-count analysis of the infrared 2MASS survey is presented, in order to
identify  faint  stream-like  structures   within  the  halo  of  the  Milky
Way. Selecting stars  with colours consistent with M-giant  stars, we find a
strong    over-density    of    sources    on    a    stream    with    pole
$(\ell=95^\circ,b=13^\circ)$, which corresponds to  the pole of the orbit of
the  Sagittarius dwarf  galaxy.  This  great-circle feature  of  width $\sim
12^\circ$, contains $\sim  5$\% of the late M-giants in  the Halo.  No other
stream-like  structures  are  detected  in  M-giants  in  the  2MASS  Second
Incremental Data Release (2IDR), and  in particular, we find no evidence for
a  stellar component  to  the  Magellanic Stream.   This  suggests that  the
present accretion rate  of low-mass satellites with a  luminous component is
very low, and the formation of  the luminous component of the Halo must have
been  essentially complete  before the  accretion of  the  Sagittarius dwarf
galaxy,  more than  $3\Gyr$ ago.   We also  search for  great-circle streams
using almost all high-latitude ($|b|>30^\circ$) sources in the 2IDR dataset.
No narrow  great-circle streams of width  $0.5^\circ$--$2^\circ$ were found,
though we were  only sensitive to relatively nearby  ($<17\kpc$) remnants of
massive ($10^6\msun$) globular clusters.  If the Galactic potential is close
to  being  spherical, as  some  recent  observations  suggest, the  lack  of
observed great-circle streams is consistent with the presence of dark matter
substructures  in the  Halo.   Although alternative  explanations cannot  be
ruled out  from our  analysis of the  2IDR dataset, future  experiments with
better statistics  have the potential to  reveal the heating  effect of dark
matter substructure on stellar streams.
\end{abstract}


\section{Introduction}

According to our current understanding  of the formation of galaxies, galaxy
halos build up  over time by the continued accretion  of smaller galactic or
sub-galactic structures  \citep{cole}.  Of  the theories that  quantify this
formation process,  the Cold  Dark Matter theory  gives at present  the best
explanation of available data, but there are several inconsistencies in this
picture that remain to be  explained. Possibly the most troublesome of these
is the inconsistency between the  predicted and observed density profiles of
galaxy halos. CDM predicts  strongly centrally-peaked ($\rho \propto r^{-1}$
to  $\rho \propto r^{-1.5}$)  mass distributions  \citep{navarro, moore99a},
contrary  to what  is inferred  from  the slowly-rising  rotation curves  of
highly  dark  matter  dominated  dwarf disk  galaxies  \citep{cote,  deblok,
marchesini}.   Another  problem is  the  apparent  disagreement between  the
expected  number of  sub-structures in  CDM  halos compared  to the  visible
substructures  in  the  halos  of  real  galaxies  \citep{klypin,  moore99b,
moore01}. It appears  that a proper treatment of the  ionizing effect of the
intergalactic UV  background, which disperses the baryonic  component of low
mass proto-galaxies  into the intergalactic medium, can  explain the absence
of   copious  low-mass   substructures  in   galaxy   halos  \citep{bullock,
somerville, tully}.   There is also  some concern that  current cosmological
simulations  are  of  insufficient   resolution:  for  instance,  one  needs
excellent numerical  resolution to  properly resolve the  wakes in  the dark
matter  distribution  which  give   rise  to  dynamical  friction.   If  the
resolution is not high enough,  dynamical friction will be artificially low,
and  structures  will  take longer  to  merge  than  they would  in  reality
\citep{kampen}.

Further observations are  required to help resolve these  issues with CDM or
to guide galaxy  formation theory. Data that constrain  the structure of the
luminous and  dark matter  in the inner  regions of galaxies  are relatively
easy to obtain, but in those  regions the dynamical times are short and much
of  the detailed  information about  the formation  of the  galaxy  has been
phased-mixed and  is now  largely lost.  In  the outer regions  of galaxies,
typical dynamical  mixing timescales are long, and  so valuable information
on the formation of the galaxy can  be recovered, if only we can find tracer
material or dynamical probes.  This is, of course, the challenge of studying
the  halos of  galaxies:  their  very low  surface  brightness renders  them
unobservable, in general, to present instrumentation.

One solution to this problem is to study the halo of the Milky Way and other
nearby galaxies, where the halo stars are resolved. Radial velocities can be
obtained  in this situation  to complement  the starcounts  data, allowing
detailed dynamical modeling of the  stellar population. In this way one can
study the structure  and substructure of the underlying  dark matter even if
it is not  traced by stars. The nearby universe therefore  gives us a unique
and  more complete  view  of the  dark  matter, which  we  cannot gain  from
statistical studies of more distant systems. The richness of the information
in  this region  will be  even  more striking  after the  completion of  the
astrometric satellite missions GAIA and  SIM: we will then also have precise
measurements of the transverse motion of a huge sample of Galactic and Local
Group stars.

In  a   companion  paper   (\citealt{ibata02};  Paper~1),  we   discuss  the
possibility of using narrow tidal  streams from defunct globular clusters to
ascertain the lumpiness of galactic halos.   We show that the heating due to
repeated close encounters with the dense dark matter clumps predicted by CDM
and  other structure  formation theories  spread an  otherwise  narrow tidal
stream into a  coarse band. Individual streams from  the expected population
of tidally destroyed globular clusters should be readily detectable with the
GAIA all-sky survey and the effect of the broadening of the globular-cluster
tidal stream should be a strong  effect if CDM predictions are correct.  The
GAIA dataset will  be by far the most efficient way  to reveal these ancient
tidal  streams,  since there  will  be  proper  motion measurements,  radial
velocities as  well as a  good spectral energy distribution  from multi-band
photometry.   The  kinematic  data  will  allow  selection  of  stars  in  a
5-dimensional phase-space  (perhaps even in 6-dimensions  for nearer streams
for which parallax measurements are possible).

However, most of the giant-branch stars of the streams that will be detected
by GAIA  have already  been cataloged in  current all-sky  surveys.  Ancient
accretions of low-mass  galaxies and globular clusters in  the outer halo of
the Milky Way  should give rise to long  stellar streams \citep{johnston96}.
These streams may be detected  by searching for long band-like overdensities
on the sky if the Halo potential is not very flattened.

In this contribution we undertake a preliminary analysis of the 2 Micron All
Sky Survey (2MASS) to investigate if  any streams can be identified from sky
position together  with simple colour-magnitude cuts  that reject foreground
contamination.   Our  two-dimensional search  is  therefore  only likely  to
recover the structures of greatest contrast over the background.

Some  tidal  streams  from  disrupting  stellar systems  have  already  been
identified.  \citet{leon} have found  low-level tidal streams around several
globular clusters,  and another very nice  example has been  revealed by the
Sloan  Digital  Sky  Survey  \citep{odenkirchen}.   These  globular  cluster
streams, however, are a most a few degrees long and it has not been possible
to trace them over  a large extent on the sky, which is  needed if we are to
be able to  probe the lumpiness of the halo  dark matter.  Large-scale tidal
streams of more  massive systems such as disrupted  dwarf galaxies have also
been  found.  In another  companion paper  \citep{ibata01a}, we  analysed the
distribution of  Carbon stars (C-stars)  in the halo  of the Milky  Way, and
found that more  than half of these are distributed along  a great circle on
the  sky.  These C-stars  trace the  tidal stream  of the  Sagittarius dwarf
galaxy,  which  has  been  disrupted  over  the course  of  its  many  close
encounters with the Milky Way.  In extragalactic systems, huge tidal streams
have been found around NGC~5907 \citep{shang} and M31 \citep{ibata01b}.

\section{Sample~1: selection of late M-giant Halo stars}

The dataset  we use  is the  Second Incremental Data  Release (2IDR)  of the
2MASS  project.~\footnote{see  \tt  http://www.ipac.caltech.edu/2mass}  This
release covers 47\%  of the sky in  a rather patchy fashion, as  can be seen
from  Figure~1, which  shows  the density  of  point-source detections  with
$K<14.3$ (the 10 sigma limit of the 2MASS photometry).  The vast majority of
the stars in Figure~1 belong to the Galactic disk and bulge; these outnumber
spheroid  stars by  several orders  of magnitude.   They are  the background
``noise''  that has  to be  removed, or  accurately modelled  out,  from the
sample.

Since the structures  we are interested in revealing  are at large distances
in the Halo, it is convenient to  remove first the disk stars from the 2MASS
sample.  To do  this we took advantage of the structure  of the $J-H$, $H-K$
colour-colour diagram.   As shown by  \citet{bessel}, dwarf and  giant stars
beyond M1 separate cleanly, the  M-giants being 0.1 to 0.3 magnitudes redder
in  $J-H$ than  M-dwarfs  at a  given  $H-K$ colour.   Selecting stars  with
$(J-H)_0  >  0.75 +  2\delta(J-H)$,  $(H-K)_0  <  1.5 -  2\delta(H-K)$,  and
$(J-H)_0 > (H-K)_0  + 0.4 + 2\delta(H-K)$ removes 99.9\% of  all of the 2IDR
2MASS point-source detections, the majority of which are nearby disk dwarfs.
(The colours  were corrected for interstellar reddening  by interpolating in
the  IRAS-DIRBE  reddening  map  of  \citet{schlegel}, using  $A_J  =  0.902
E(B-V)$, $A_H =  0.576 E(B-V)$, $A_K = 0.367  E(B-V)$).  Though the relative
fraction of dwarf  and giant stars in the disk and  spheroid may be similar,
by  selecting only  giant  stars, we  are able  to  clean much  of the  disk
contribution to the  high-latitude sky, but even with  this cleaning we find
that  the  contamination   at  20--30  degrees  from  the   Plane  is  still
sufficiently  high that  we are  restricted to  using a  sample with  $|b| >
30^\circ$.

Given a characteristic absolute magnitude of RGB M-giants of $M_K \sim -5.5$
\citep{nikolaev}, the bright limit of  $K=8.0$ of the 2MASS data corresponds
to a lower distance limit of  $5\kpc$ for the giant stars.  This distance is
more than three thick disk scale-heights ($z_h \sim 800\pc$ \citealt{reyle})
for  $b>29^\circ$.  (The  2MASS faint  limit of  $K=14.3$ corresponds  to an
upper distance limit of $\sim100\kpc$.)

We  decided   to  ignore   the  data  at   Galactic  latitudes   lower  than
$|b|=30^\circ$, as contamination  from the Galactic thick disk  begins to be
significant below  that latitude.  Strong  reddening can move dwarfs  in the
region of the colour-colour diagram  inhabited by giants, so we also excised
from our  final dataset all regions  of the sky where  the reddening deduced
from the IRAS-DIRBE map  of \citet{schlegel} exceeded $E(B-V)=0.1$.  Because
of this reddening  cut, and to maintain a homogeneously  deep map, we further
restricted  the  K-band  limit  to  $K=14.3-0.367  E(B-V)_{max}=14.26$.   At
Galactic latitudes $|b|>30^\circ$, the area  of sky which is removed by this
extinction cut  is very  small ($0.6\%$) and  is confined to  regions around
nearby  molecular clouds.   Finally, two  more regions  are cut  out  of the
dataset: a circular region of radius  $10^\circ$ around the LMC and SMC. The
resulting map  is shown in Figure~2,  which shows the  1106 point-sources on
the sky that remain after these  various cuts; the area of sky available for
analysis (see Figure~3) is $26.4$\%.

\begin{table}
\caption{Parameter cuts employed for selection of the M-giant star sample.}
\begin{tabular}{rcc}
parameter &  & value \\
\\
$(J-H)_0 - 2\delta(J-H)$           &  $>$ &  $0.75$ \\
$(H-K)_0 + 2\delta(H-K)$           &  $<$ &  $1.5 $ \\
$(J-H)_0 - (H-K)_0 - 2\delta(H-K)$ &  $>$ &  $0.4 $ \\
$K$       &  $<$ &  $14.26$    \\
$|b|$     &  $>$ &  $30^\circ$ \\
$d_{LMC}$ &  $>$ &  $10^\circ$ \\
$d_{SMC}$ &  $>$ &  $10^\circ$ \\
$A_V$     &  $<$ &  $0.1$      \\
\end{tabular}
\end{table}

\section{Pole-count analysis of sample~1}

One  way to  identify  long  band-like structures  is  through a  pole-count
analysis.  The  pole-count analysis  is very simple:  the number  of sources
situated within a band centered on a  great circle on the sky is counted up,
and assigned to  the corresponding pole.  This is  repeated for all possible
poles. In  this way, a band-like structure  on the sky becomes  two peaks in
the  pole-counts diagram (two  peaks due  to the  symmetry of  the mapping),
whereas a localised peak in counts on  the sky becomes two bands in the pole
counts  diagram. Due  to  the  complicated irregular  nature  of the  window
function, it is  more convenient to perform the  pole-counts on binned data:
to  this end we  binned up  the sky  into a  rectangular $3600  \times 1800$
grid. To correct  for the sky regions for  which we have no data,  we add in
the average  number density  of sources  into each pixel  in the  image (see
Figure~3).

The  pole-counts  analysis  is   particularly  powerful  if  the  underlying
potential in the region of the  Galaxy inhabited by the stream is spherical,
in that case the  stellar stream does not precess and it  remains on a great
circle  on  the  sky  forever.   In  a  halo  with  a  flattened  potential,
differential  precession  will  eventually  phase-mix  the  stream,  thereby
filling  the box-orbit.  In  this situation,  only strong  recent accretions
would be detected by the pole-counts technique.

Fortunately for the  pole-counts method, the outer regions  of the Milky Way
appear to have  an almost spherical potential.  The  evidence for this comes
partly from the  great-circle distribution of C-stars from  the tidal stream
of  the  Sagittarius  dwarf   galaxy  \citep{ibata01a},  and  also  from  the
distribution  of metal-poor  spheroid stars  towards $R\sim  20\kpc$  of the
inner Halo region surveyed by \citet{chiba}.

The free  parameter that we  need to choose  is the width of  the equatorial
bands.   Clearly,  narrow  bands   are  better  suited  to  identify  narrow
structures.  We chose not  to probe  band widths  below $1^\circ$;  this was
motivated by the simulations of  Paper~1, where model globular clusters gave
rise to  stream widths  of $\simgt 2\kpc$  (equivalent to $\sim  1^\circ$ at
$100\kpc$). The analysis was  repeated with band widths $1^\circ, 1.5^\circ,
2^\circ, \dots,  20^\circ$, and peaks  $>3\sigma$ above the  background were
identified in  each polecounts  image.  The result  of the pole-count  for a
band width  of $12.0^\circ$ (which gives  the highest S/N peak)  is shown in
Figure~4. Of our Halo sample of 1106 sources, $15$\% are located within this
$12^\circ$ band,  compared to $10$\% that  would be expected  from a Poisson
distribution. How significant is this peak given the strange window function
of  the  2IDR  data?   To  answer  this question,  we  have  performed  1000
Monte-Carlo simulations,  drawing in each  case 1106 random points  over the
sky within the data regions displayed in Figure~3. The analysis was repeated
in exactly  the same way as  for the real  data, searching for peaks  in the
pole-counts diagram.  In the  1000 simulated datasets, the highest deviation
was a $12.9$\% peak, found at  an $(\ell,b)$ location far from the Sgr pole.
Interpreting this  maximum random deviation  as a $3.2\sigma$  peak, implies
that the  peak in  the pole-counts distribution  detected in Figure~4  has a
significance of $\sim 5.5\sigma$ and is therefore clearly real.

\section{Sample~2: almost no restrictions}

In selecting only  late M-giant stars from the  2MASS dataset, we previously
retained only a small fraction of  the full sample of Halo stars observed by
that  survey.  To  try  to overcome  this  handicap, we  undertook a  second
analysis, this time with a much broader colour selection, aimed at including
K-stars and early M-stars as well  as late M-stars selected for the previous
sample.  The selection criteria are  simple colour limits: $-0.1 < (H-K)_0 <
0.5$, $0.1 < (J-H)_0  < 1.0$; a faint magnitude limit: $K_0  < 14.26$; and a
selection of  only low  extinction sky $E(B-V)  < 0.1$.  Selecting  from the
high latitude sky  ($|b|>30^\circ$) in the 2IDR with  these criteria gives a
much larger sample of $5.9\times10^6$ point sources. The challenge now is to
be able to  extract a stream population with at most  a few thousand sources
from such a numerous foreground population.

To  successfully identify  streams with  the polecounts  method, we  need to
model out these contaminants.  However, the distribution of Galactic sources
is not  uniform, and it is  only approximately symmetric  about the Galactic
major and minor  axes. We found that  it was necessary to create  a model of
the smooth Galactic foreground which was  able to follow the variations on a
scale of  $\sim 10^\circ$; smoother  models give rise to  unacceptably large
residuals which dominate the subsequent polecounts.  The complex 2IDR window
function also makes it difficult  to fit an accurate model: e.g., polynomial
fits of high enough order  to follow the $\sim 10^\circ$ Galactic variations
tend to blow  up in the unconstrained regions,  leaving systematic residuals
around the holes.

By  trial and  error, we  managed to  construct an  acceptable model  of the
Galaxy  by smoothing the  data. However,  to avoid  edge effects  around the
holes  in the  2IDR  dataset, the  holes  needed to  be  filled before  this
smoothing operation. The empty  regions were filled preferentially with data
from the symmetric  pixel reflected about $l=0$, otherwise  by data from the
pixel reflected  about $b=0$ or the  pixel reflected about  $l=0$ and $b=0$.
After this  operation, 8.6\% of the  sky still lacked  data; these remaining
holes were  filled with a  Poisson random deviate  drawn from a  third order
polynomial  fit to each  longitude line  separately.  The  resulting all-sky
image was smoothed  with a $5.1^\circ$ bi-linear kernel  (taking the average
of counts within $\pm 4\sigma$ of the median).

An image  of the  difference between the  original dataset and  the smoothed
model was  created and  the window function  was reinserted into  the image.
This difference  image is now flat  to $\sim 2$\%. The  difference image was
smoothed again  with a  $5.1^\circ$ bi-linear kernel,  but this  time, those
regions with  data on either side  of a hole  were (for the purposes  of the
smoothing) made  contiguous.  After removing this smoothed  model, the final
dataset is flat to $\sim 0.5$\%.

This image of the residuals of the 2IDR from a smoothed model is appropriate
for searching for narrow great-circle streams of width up to $\sim 2^\circ$.
Polecounts diagrams were computed  for bands of $2.0^\circ$, $1.0^\circ$ and
$0.5^\circ$. Within  a $2^\circ$ wide band  (1.7\% of the  sky) $10^5$ stars
are expected,  on average, with  $1\sigma$ variations of $320$  counts.  The
corresponding  polecounts diagram  had an  R.M.S.  dispersion  $450$ counts,
close to this  Poisson limit.  Since $2^\circ$ wide bands  will give rise to
$\sim 2^\circ$  wide peaks  in the polecounts  image, there are  $\sim 5000$
``meta-pixels'' in  that image, so  $>4\sigma$ peaks are required  to reject
most false detections due to random noise.  There is a single peak with more
than 1800  counts, which appears  as an arc  in the polecounts  diagram near
$(\ell=0^\circ,b=0^\circ)$.  The  feature is  clearly not real,  however, as
the corresponding  equatorial band passes through  a long narrow  gap in the
window  function  that  skims  the  outer  bulge  of  the  Milky  Way  (near
$b=-30^\circ$).   The polecounts  peak disappears  if we  remove  this small
region  of sky from  the analysis,  so the  peak is  clearly an  artifact of
insufficient accuracy in the smooth Galaxy model in that region.

With very narrow $0.5^\circ$ bands,  the R.M.S. dispersion of the polecounts
image  was $215$  counts,  so streams  with a  few  as 860  stars could,  in
principle, have been detected. No such peaks were detected, however.

\section{Discussion}

\subsection{Wide streams from disrupted dwarf galaxies}

The 2MASS 2IDR data reveal a strong overdensity along the great circle track
of the Sgr dwarf galaxy. Indeed, the pole-counts diagram of Figure~4 is very
similar to that  found by \citet{ibata01a} from Halo  C-stars.  M-giants are
potentially older stars than C-stars, and  as such they can be useful in the
context of  analysing the Sgr stream,  allowing us to probe  the stars which
were stripped from the dwarf galaxy at earlier stages in the disruption.  To
constrain this possibility, the radial  velocities of the M-giants needed to
be measured.

For  bands of  width  $<2^\circ$, the  Sgr  polecounts peak  breaks up  into
several  lower $S/N  \sim  3$ peaks,  presumably  because the  orbit is  not
precisely on a great circle.  This  could potentially be due to a flattening
of  the Galactic potential  (the Galactic  disk clearly  provides non-radial
forces); an analysis  of this effect could be attempted  with the full 2MASS
dataset.

It is surprising  not to have found stream-like structures  in the Halo from
other  dwarf  satellites,  such  as  the Magellanic  Clouds.   Our  previous
polecount  analysis of  Halo C-stars  provided  a hint  that the  Magellanic
Stream had been  detected.  Despite the better statistics,  we have not been
able to confirm that tentative  detection. It is possible that this reflects
a difference  in stellar populations between the  Sagittarius and Magellanic
streams, the  latter possibly  being richer in  C-stars due to  younger age.
Another explanation  is that the current  sky coverage excludes  most of the
region in and around the Magellanic  Clouds, this makes the detection of the
Magellanic  Stream  unlikely with  the  current  dataset  if the  stars  are
distributed in a similar fashion to the gas in the Magellanic Stream.

Our models of  the disruption of the Sgr dwarf suggest  that the streams are
very  long  lived, remaining  coherent  for a  large  fraction  of a  Hubble
time. However, differential precession can  cause a stream to smear out over
the  sky, until  the box  orbit is  filled up  entirely.   Two possibilities
present themselves.   Either the Halo  is fairly flattened, contrary  to the
results of \citet{ibata01a} and  \cite{chiba}.  In this case streams precess
quickly, and  pole-counts give us a  snapshot of accretions  that have taken
place only very recently, on a  timescale of $\sim 1\Gyr$ or less (depending
on the orbit).  If the Halo  is close to being spherical, polecounts give us
access  to  much  more  ancient  accretion  events.   Without  a  conclusive
measurement  of  the  flattening  of   the  Halo,  we  cannot  place  strong
constraints on the merging rate onto the Milky Way.

However,  assuming that the  M-giant stellar  population in  the Sagittarius
stream is representative of other accretions, it appears that there has been
no other  more massive accretion than  the Sgr dwarf  ($M_{Sgr} \sim 5\times
10^8\msun$) since that galaxy was  accreted. The timescale for the accretion
of the Sagittarius dwarf must be long, of order several Gyrs, as it takes of
order $3$--$5\Gyr$  in the  simulations \citep{ibata98, ibata01a}  for tidal
tails to wrap around the Milky Way and appear as a band on the sky.

\subsection{Narrow streams from disrupted globular clusters}

As we  show in  Paper~1, the structures  of particular interest  are stellar
streams from  low mass  systems such as  globular clusters. These  give fine
structures  in  phase-space  which  easily  become  dispersed  if  there  is
significant substructure in the Halo.  No large scale stream other than that
of the Sgr dwarf was identified in the sample of late M-giant stars from the
2MASS 2IDR dataset above the $3\sigma$ level.

With our late  M-giant sample, the $3\sigma$ detection  level corresponds to
only 9 stars at the finest band resolution adopted ($1^\circ$).  The lack of
detections reflect the fact that our M-giant selection criteria, which aimed
to get rid  of disk dwarfs, are  very restrictive.  A vast number  of Halo K
(and  early M)  giants were  removed in  this process  leaving only  a small
number of Halo  stars in the final sample.   Applying our selection criteria
to  the 2MASS  data  of  the massive  ($10^6\msun$)  globular cluster  Omega
Centauri (distance $5.1\kpc$)  yields a total of $3$  stars outside two core
radii, or 5.8~stars correcting for the lack of data in the saturated central
regions of the  cluster.  Only 1 in 2000 stars above  the magnitude limit of
$K=14.3$ are kept after the  colour-cut (though the stars that are retained
are very luminous, and could be detected out to beyond $100\kpc$).  So it is
not surprising that we have not  found globular cluster streams in the first
sample, as  the tidal stream of  a completely disrupted  globular cluster of
the size  of Omega  Centauri would have  insufficient late M-giant  stars to
bring the contrast of a polecounts peak above the $3\sigma$ detection level.

The inclusion of Halo K-giants and  early M-giants is needed to obtain high
enough $S/N$ to  make a useful statement about  dark matter clumpiness.  Our
second  analysis,  with  a   sample  selected  with  practically  no  colour
restrictions, aimed  to improve the statistics for  narrow streams. However,
no streams were  detected above the $4\sigma$ level, even  though we had the
capability of  identifying narrow  streams with as  few as $\sim  860$ stars
with  the narrowest  band  we probed  ($0.5^\circ$).  For comparison,  Omega
Centauri would  provide a stream of $\sim  3450$ stars in 2MASS,  if it were
located at  a distance of  $17\kpc$, though the  $25$\% sky coverage  of the
$|b|>30^\circ$ region of the 2IDR  dataset reduces this number to $\sim 860$
stars.

The detection  of a dynamically cold  thin stellar stream in  the 2IDR would
have called  into question the  existence of dark matter  substructure.  Our
null detection  is much less clear  to interpret, however.  We  can rule out
the existence of  only very massive globular cluster  streams (as massive as
Omega  Centauri),  that are  relatively  nearby  ($<17\kpc$),  and that  are
confined to  very narrow  great circle bands  ($<0.5^\circ$). This  may mean
that massive globular  clusters are rare, or that  the Galactic potential is
significantly flattened  at these radii,  or that the potential  has evolved
substantially since the disruption of these stellar systems.

Future experiments will  help solve this question.  By  reanalysing the full
2MASS  dataset, a $\sim  3$ times  larger area  of sky  can be  surveyed for
streams  (not $\sim 4$  times larger,  since low  latitude disk  fields will
likely be too complex  to model). With that data in hand  it may be possible
to improve the stream detection  limits by an order of magnitude, especially
if the foreground Galactic component  can be more accurately modelled in the
absence  of the  complex  2IDR window  function  (which we  believe was  the
primary obstacle  in not reaching the  Poisson noise limit  in the polecount
analysis).    Other   purely    photometric   experiments,   such   as   the
VISTA~\footnote{see    \tt    http://www.vista.ac.uk}    survey    or    the
PRIME~\footnote{see  \tt http://prime.pha.jhu.edu/primemission.htm} mission,
will also greatly improve stream detection limits due to improved statistics
resulting from  a much greater  photometric depth than 2MASS.   Larger gains
will come from  the addition of astrometric information.   A first step will
be to  use proper  motions derived from  photographic plates  (provided, for
example          by           the          GSC~II~\footnote{see          \tt
http://www-gsss.stsci.edu/gsc/gsc2/GSC2home.htm}), with which it will become
possible  to obtain a  cleaner Halo  sample by  requiring that  distant Halo
stars  have  proper motions  consistent  with  zero.   The analysis  of  the
datasets mentioned  until now can  provide information on Halo  streams only
for the  restricted case  that the Galactic  potential is  nearly spherical,
when  streams  are spatially-confined  bands  on  the  sky.  The  advent  of
GAIA~\citep{perryman}, with its ability to accurately measure proper motions
even for distant Halo stars,  will permit the identification of streams even
if the Galactic potential is substantially flattened, as the streams will be
seen as phase-space clumps.

\section{Conclusions}

A pole-count analysis of Halo  M-giants stars in the Second Incremental Data
Release of  the 2MASS  dataset reveals  a strong stream  of stars  along the
orbit of the Sagittarius dwarf galaxy. The pole of the Sagittarius stream is
located at exactly  the position predicted by our  previous analysis of Halo
C-stars  \citep{ibata01a},  confirming  the  existence of  this  large-scale
stellar  stream.  This  stream  appears fairly  narrow,  $\sim 12^\circ$  in
M-giants.  The alignment of the narrow stream along a great circle rules out
the possibility  that the  Halo mass distribution  is highly  flattenned, as
shown  in \citet{ibata01a}.   Future radial  velocity measurements  of these
M-stars will easily allow us  to discriminate stream from Halo or foreground
stars. Our  conclusion regarding the shape  of the dark matter  Halo will be
strengthed if  the stream stars are  found to be distributed  all around the
great circle track.

Our main motive in undertaking this study was to identify kinematically cold
streams from low mass systems such as globular clusters, which could be used
to  trace the substructure  of the  dark matter  Halo. However,  no globular
cluster streams  were identified in  the sample of Halo  late-type M-giants,
though we argue  that this is due to insufficient  statistics in the sample.
Neither were any streams detected in  the second sample that contained K and
M-giants (as  well as  numerous foreground stars).   In this  second sample,
nearby ($<17\kpc$) globular  cluster streams, of mass similar  to that Omega
Centauri, and  that are  closely confined to  a great  circle ($<0.5^\circ$)
could  have been  detected. These  stringent conditions  greatly  reduce the
chances of  detection of any Halo streams.  In reality, even if  the Halo is
spherical,  the flattened  potential of  the Galactic  disk will  cause some
precession, especially to nearby streams.  Also, the Sun's displacement away
from  the center  of the  Galactic potential  means that  only  very distant
streams can be located precisely on great circles.

Better  statistics are  needed to  allow the  detection of  globular cluster
streams on wider great-circle bands, and deeper datasets are needed to probe
more distant orbits.  It will certainly be possible to improve this analysis
in the  near future with the full  sky 2MASS catalog, and  better dwarf star
rejection  can be  provided by  proper  motion selection  (derived from  the
GSC-II catalog, for  instance).  In the longer term,  the next generation of
IR  surveys (e.g.,  VISTA, PRIME),  and the  next generation  of astrometric
surveys (e.g.,  DIVA, GAIA)  will greatly improve  the number of  faint Halo
sources and allow  us to identify individual streams  from their photometric
and kinematic signatures.

\newcommand{\mnras}{MNRAS}
\newcommand{\nat}{Nature}
\newcommand{\araa}{ARAA}
\newcommand{\aj}{AJ}
\newcommand{\apj}{ApJ}
\newcommand{\apjl}{ApJ}
\newcommand{\apjs}{ApJSupp}
\newcommand{\aap}{A\&A}
\newcommand{\aaps}{A\&ASupp}
\newcommand{\pasp}{PASP}

\onecolumn

\begin{figure}
\ifthenelse{\UseFigs=1}{
\includegraphics[angle=270,width=15cm]{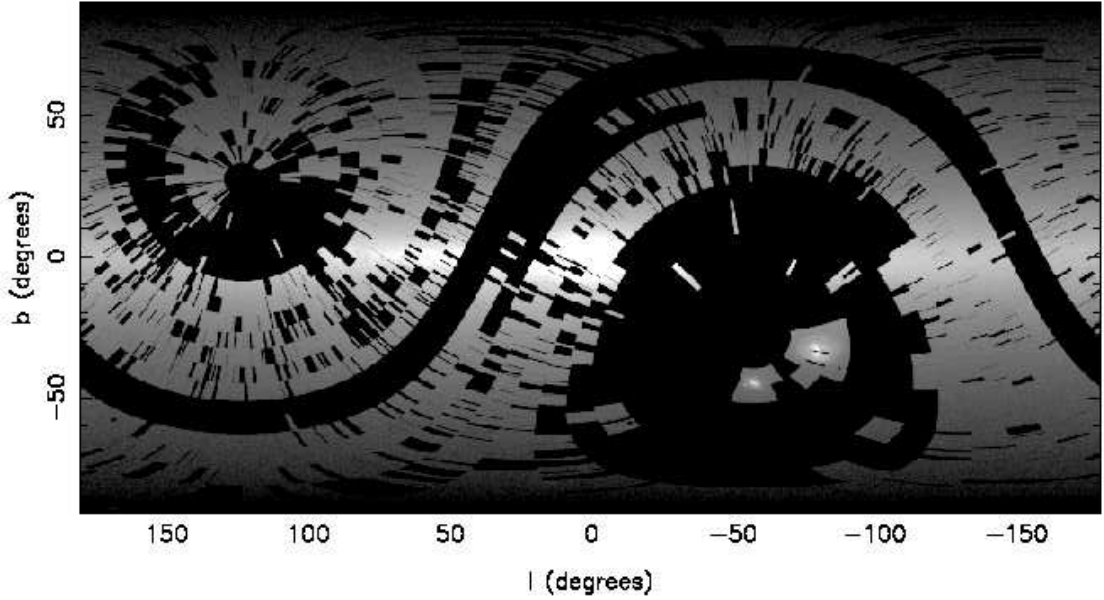}}{}
\caption{The distribution of sources over the sky in the 2IDR dataset of the
2MASS  project.  Each  $0.1^\circ  \times 0.1^\circ$  pixel (in  rectangular
coordinates) in the image displays  logarithmically the number of sources in
the corresponding  region that have  $K<14.3$ (the $\sim 10\sigma$  limit of
the photometry).}
\end{figure}

\begin{figure}
\ifthenelse{\UseFigs=1}{
\includegraphics[angle=270,width=15cm]{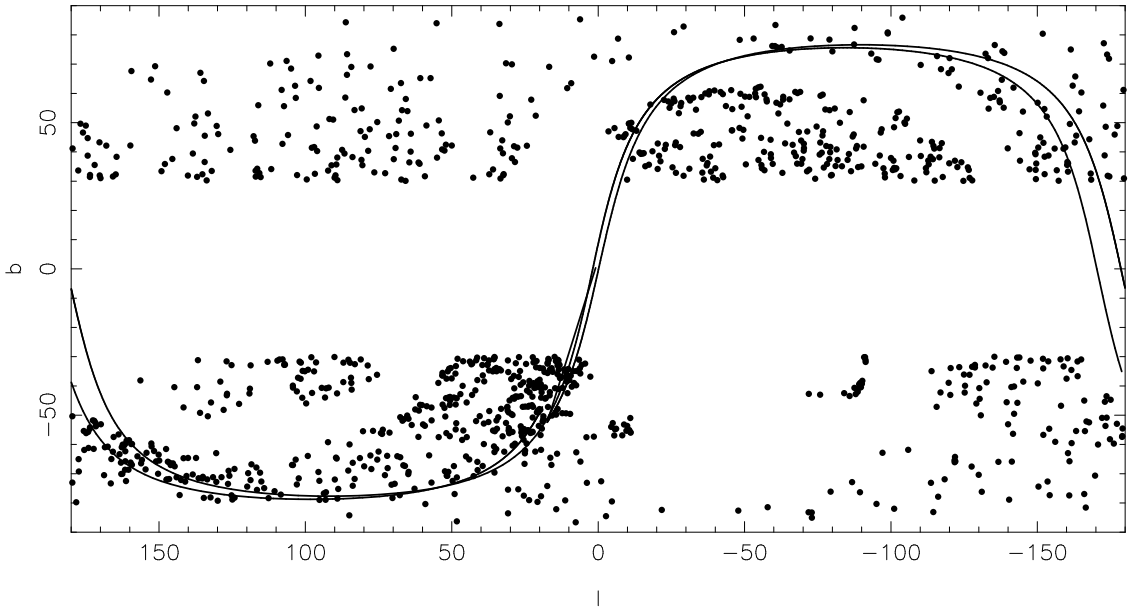}}{}      
\caption{The distribution  in Galactic coordinates of the  1106 late M-giant
Halo star candidates selected for analysis. The expected orbital track of
the Sgr dwarf galaxy (from \citealt{ibata01a}) has been overlaid.}
\end{figure}

\begin{figure}
\ifthenelse{\UseFigs=1}{
\includegraphics[angle=270,width=15cm]{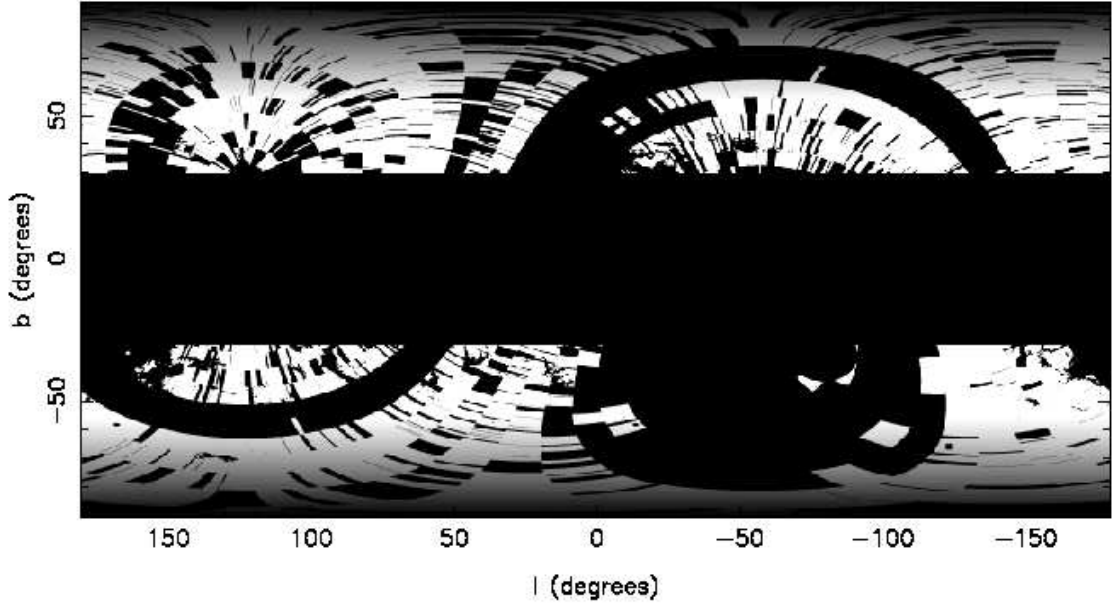}}{}
\caption{The  sky  coverage  for  the  present analysis.   The  black  areas
indicate the unobserved or unused regions  of sky, while gray areas show the
regions from which  data were obtained, with a  shading corresponding to the
area subtended per  pixel ($0.01$ square degrees on  the Equator, decreasing
to the poles).}
\end{figure}

\begin{figure}
\ifthenelse{\UseFigs=1}{
\includegraphics[angle=270,width=15cm]{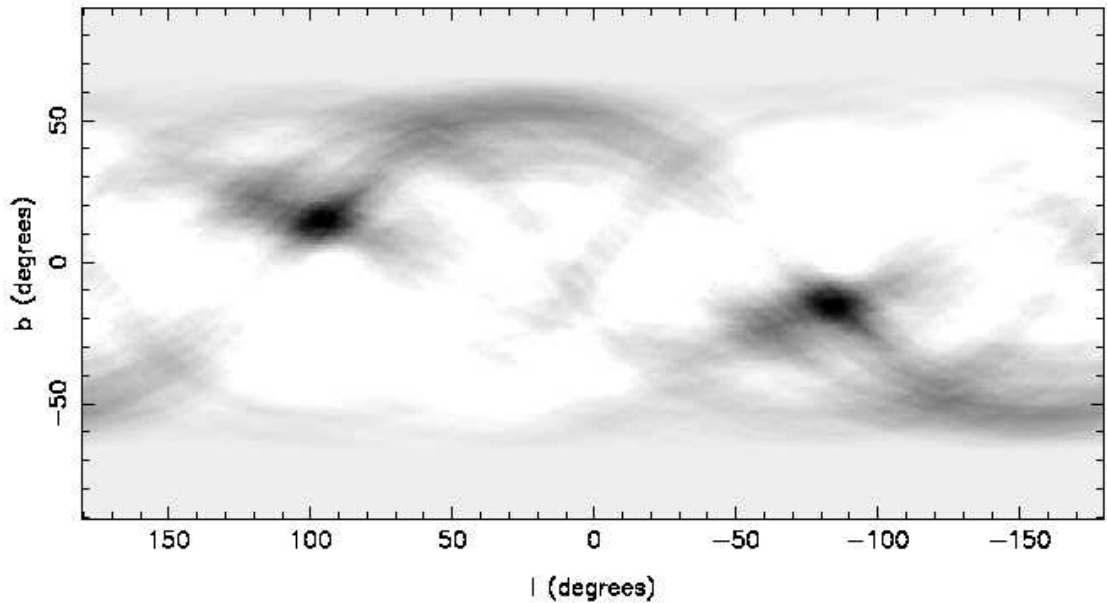}}{} 
\caption{The  pole-counts diagram of  our sample  of Halo  late-type M-giant
stars,  for   a  band   of  width  $12.0^\circ$.    The  strong   peak  near
($\ell=95^\circ,b=13^\circ$)   reflects  15\%  of   the  sample,   an  $\sim
5.5\sigma$  overdensity. Its  location,  at the  pole  of the  orbit of  the
Sagittarius dwarf galaxy, implies that  we have detected the evolved stellar
populations  of  the tidal  stream  of  that  galaxy, which  was  previously
detected in C-stars. This polecounts  diagram has a similar structure to the
polecounts   diagram   constructed    from   Halo   C-stars   (Figure~5   of
\citealt{ibata01a}), with the  difference that there is no  evidence for the
Magellanic stream in the present M-giant sample. This difference is likely a
consequence of  the lack of sky  coverage around the South  Ecliptic Pole in
the 2IDR dataset.}
\end{figure}

\end{document}